\documentclass[nature,superscriptaddress,dvipsnames,twocolumn]{revtex4-1}

\usepackage{fancyhdr} 
\usepackage{layout}
\usepackage{color}
\usepackage{amsmath}
\usepackage{tikz}
\usepackage{xcolor}
\usetikzlibrary{arrows}

\usetikzlibrary[topaths]

\usepackage{siunitx}
\usepackage{eurosym}
\usepackage{amsfonts}
\usepackage{amsmath}
\usepackage{amsthm}
\usepackage{amscd}
\usepackage{amssymb}
\usepackage{amsxtra}
\usepackage{bm}
\usepackage{bbm}         
\usepackage{exscale}
\usepackage{graphics,color}
\usepackage{latexsym}
\usepackage{enumerate}
\usepackage{dcolumn}      
\usepackage{graphics}
\usepackage{epstopdf}
\usepackage[dvips]{epsfig}
\usepackage{graphicx}
\usepackage{hyperref}
\usepackage{tabularx}
\usepackage[authormarkup=none]{changes}
\usepackage[caption=false]{subfig}
\usepackage{lineno}
\usepackage[utf8]{inputenc} 
\usepackage{eurosym}

\usepackage{setspace}

\hypersetup{
    colorlinks,%
    citecolor=black,%
    filecolor=black,%
    linkcolor=black,%
    urlcolor=black
}

\newcommand{\beq}{\begin{equation}}
\newcommand{\eeq}{\end{equation}}

\newcommand{\bdf}{\begin{defn}}
\newcommand{\edf}{\end{defn}}

\definechangesauthor[name={Rob}, color={orange}]{rob}

\newtheorem{defn}{Definition}

\newcount\mycount

\begin{document}



\title{\textbf{SATELLITE QUANTUM COMMUNICATIONS WHEN MAN-IN-THE-MIDDLE ATTACKS ARE EXCLUDED}}
\author{Tom Vergoossen}
\affiliation{Centre for Quantum Technologies, National University of Singapore}
\thanks{S15, 3 Science Drive 2, Singapore 117543}

\email{cqtalej@nus.edu.sg}

\author{Robert Bedington}
\affiliation{Centre for Quantum Technologies, National University of Singapore}
\author{James A Grieve}
\affiliation{Centre for Quantum Technologies, National University of Singapore}
\author{Alexander Ling}
\affiliation{Centre for Quantum Technologies, National University of Singapore}
\affiliation{Department of Physics, National University of Singapore}

\date{\today}

\begin{abstract}

An application of quantum communications is the transmission of qubits to create shared symmetric encryption keys in a process called Quantum Key Distribution (QKD). Contrary to public-private key encryption, symmetric encryption is safe from (quantum) computing attacks, i.e. it provides forward security and is thus attractive for secure communications. In this paper we argue that for free-space quantum communications, especially with satellites, if one assumes that man-in-the-middle attacks can be detected by classical channel monitoring techniques, simplified quantum communications protocols and hardware systems can be implemented that offer improved key rates. We term these protocols photon key distribution (PKD) to differentiate them from the standard QKD protocols. We identify three types of photon sources and calculate asymptotic secret key rates for PKD protocols and compare them to their QKD counterparts. Results show that PKD protocols have roughly a factor of two higher rates as only one measurement basis is used and due to the relaxed security assumptions can establish keys at very high losses whereas in QKD the privacy amplification process becomes prohibitive.

Key words: quantum communication; quantum cryptography; quantum key distribution; satellite QKD; access control; threat model; photon key distribution; PKD
\end{abstract}

\begin{titlepage}
\maketitle
\clearpage

\end{titlepage}



\section{Introduction}


Cryptographic key distribution is a major application of quantum communication. Such schemes typically use measurements of quantum states of photons shared between two remote parties to allow both sides to derive shared entropy that can be quantitatively assessed to be private. This shared entropy may be used as keying material for use as one time pads \cite{Vernam1926} or as seed keys for symmetric encryption \cite{NISTFIPS2001}. Quantum Key Distribution relies on transmission of single photons so many photons must be distributed to generate a key over lossy channels. For shorter distances and metropolitan regions photons can be distributed between parties using optical fibres, but for global distances satellite-based nodes become more practical \cite{Bedington2017}. For quantum communication from satellites, and other moving platforms, photons are distributed using free space optics (FSO). 

Quantum key distribution schemes \cite{Gisin2002,Diamanti2016a} have extremely strong security guarantees due to minimal assumptions on the capabilities of the technology available to potential eavesdroppers---essentially any attack permitted by the laws of physics is deemed possible. Although typically the two communicating parties must trust the QKD hardware within their control, they need not trust the optical channel between them because they can detect any attempt at eavesdropping and any man-in-the-middle attacks on the channel using statistical tests that are inherent to the QKD process. These tests require a fraction of the received photons to be discarded such that they cannot be used in the final keying material. This discarding arises in processes such as basis reconciliation, parameter estimation and privacy amplification. In many situations these discarded photons and slower key rates are a necessity for security e.g. in a metropolitan environment where QKD is performed over optical fibres which pass through many ducts and underground passages where, in principle, eavesdropping can take place. 

For the case of satellite QKD (and other FSO delivery methods), where there is a direct line of sight, these measures seem to be harder to justify. To compromise the security of the link, other than disrupting it through denial-of-service attacks, an adversary would have to act as a man-in-the-middle. We argue that performing this attack for an optical link between a low earth orbit (LEO) satellite and ground station would be physically possible, but technically not feasible for most adversaries. It requires intercepting and re-sending a beam that is only a few metres across and rapidly tracking across the sky without being detected. Practical satellite-ground links employ dual tracking beacons to establish the required high accuracy pointing link, effectively providing a channel monitoring system. Additional hardware may be deployed to monitor the channel in other wavelengths, e.g. radar systems or thermal imaging cameras. Furthermore, space situational awareness has led to publicly accessible catalogues that provide ephemeris data of orbiting space objects which are frequently updated, making it possible to assess if any (publicly known) satellites can threaten the link in space. If users fear that their adversaries could have technologies which can covertly intercept the link they could of course use QKD; other users can relax their threat assumptions. In light of the above we thus focus only on passive eavesdroppers, attackers which cannot be detected by channel monitoring techniques. As we will show, using this updated threat model simplifies the key delivery protocol.

In this paper we term the simplified quantum communication schemes that address this scenario as `photon key distribution' (PKD) so as to separate them from conventional QKD protocols. 
Some more recent QKD schemes, where the security proofs may not yet be completely agreed, may also prove to be somewhere on the PKD to QKD spectrum depending on the security assumptions they require. While most QKD schemes use two bases for encoding, in our PKD schemes only a single encoding basis is used since mixed bases are primarily a mechanism to detect man-in-the-middle attacks. This avoids the basis reconciliation sifting stage and so is more efficient in its use of photons.

Sasaki et al. have previously discussed the opportunity for simplified cryptography schemes that fall between QKD and laser communications \cite{Sasaki2017}. As they explain QKD (and PKD) fit within the wider field of physical layer security methods. These encompass such techniques as wiretap channels \cite{Wyner1975} and radio frequency equivalents \cite{Poor2017}, to laser-based methods that actively quantify the amount of information an eavesdropper can receive \cite{Fujiwara2018}. Our exploration of PKD here builds on the ideas put forward by Sasaki et al. 

In this paper we look at the asymptotic secret key rates at high losses (i.e long distance free space links) for three QKD schemes using three different kinds of transmitter hardware, and compare this to three simplified PKD schemes that use similar hardware. The simplified schemes are primarily targeted at satellite applications and all assume that a covert man-in-the-middle attack has been assessed to be non-existent. As expected, they are all found to remain effective at higher losses than their QKD counterparts.

\section{Results}

Table~1 shows a summary of the hardware types and corresponding QKD and PKD methods considered. Figure~1 shows a comparison of the quantum bit error rate (QBER) and bits per pulse achieved with these methods at increasing distances (losses). In this way the results can be presented independently of the pulse rate of any particular hardware setup. Although we use the term 'pulse' here for straightforwardness of comparison, the SPDC sources we consider are operated by a continuous-wave pump, i.e. not pulsed, and produce photon pairs on a stochastic basis so `detection events' would be the more appropriate term. Figures~1 (a) and (b) show that for QKD protocols (BB84, BBM92) there is a sharp drop off in key rate as losses increase. At high losses more key has to be discarded in privacy amplification than is available so the secure key rate drops to zero. This does not happen in the PKD equivalents as it is assumed that there are no active man-in-the-middle attacks, and so privacy amplification is only required for multi-photon pulses. 
\begin{table*}
\label{tab:schemes}
\caption{QKD and PKD methods modelled for different hardware implementations.\\}

\begin{tabular}{ccc}

 & & \\
\textbf{Photon Source}	& \textbf{QKD protocol}	& \textbf{PKD encoding (example)}\\

\toprule
 & & \\

Weak coherent pulse (WCP)	                       					& Decoy state BB84      			& Pulse position modulation (PPM)\\
\\\multicolumn{1}{c}{\begin{tabular}[c]{@{}c@{}}Spontaneous parametric \\ downconversion (SPDC) pairs 	\end{tabular}} & \multicolumn{1}{c}{\begin{tabular}[c]{@{}c@{}}BBM92\\ (entanglement-based BB84)\end{tabular}}&  \multicolumn{1}{c}{\begin{tabular}[c]{@{}c@{}}Heralded left or right-handed \\ polarised photons\end{tabular}}\\

\\
Ideal single photon source (SPS)                       		& Single photon BB84	    		& \multicolumn{1}{c}{\begin{tabular}[c]{@{}c@{}}Left or right-handed \\ polarised photons\end{tabular}}\\
 & & \\

\hline

\end{tabular}
\end{table*}




\begin{figure*}[t!p]
\subfloat[]{
            \centering
            \includegraphics[width=.5\textwidth]{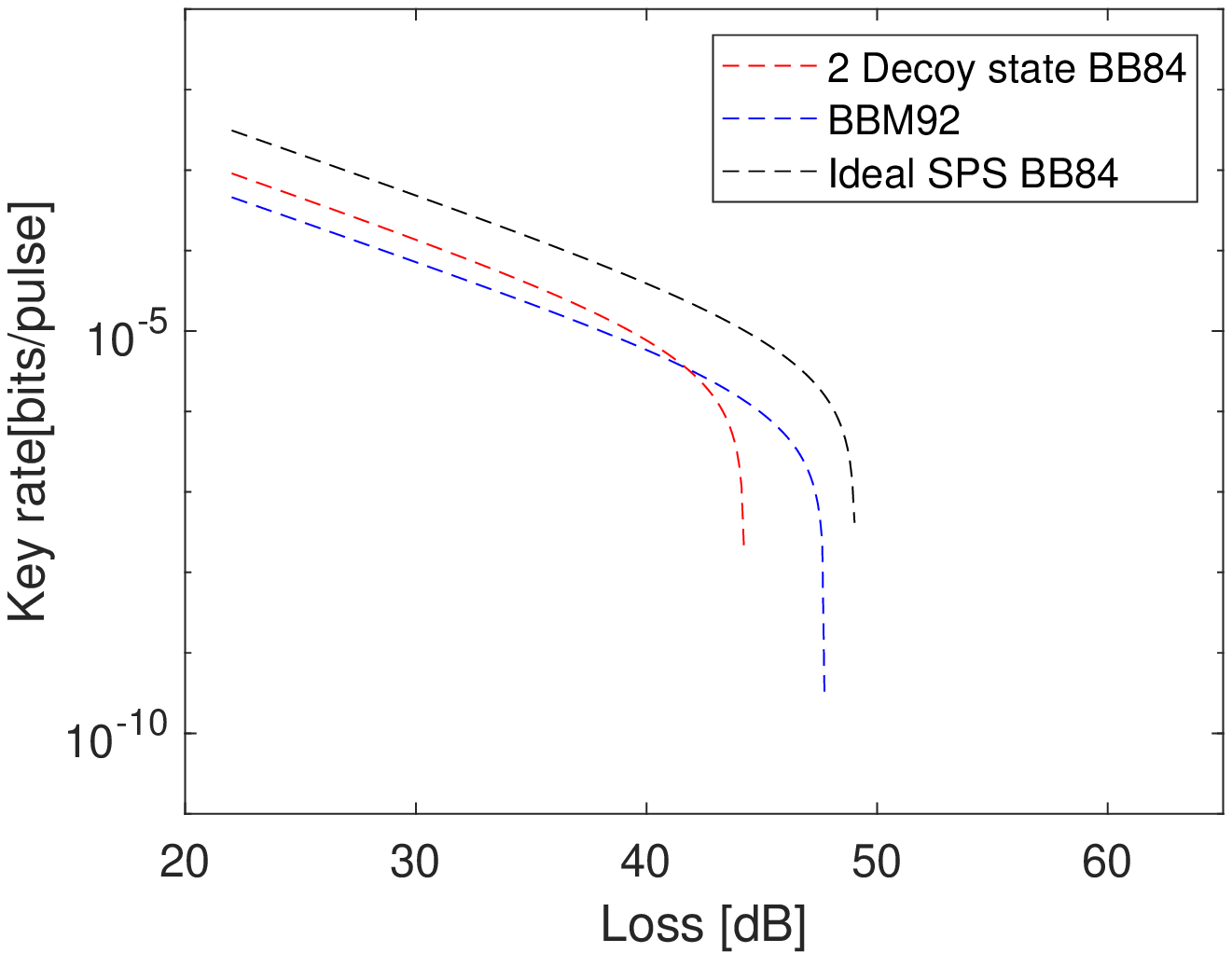}
            \label{fig:qkd_key}
        }
\subfloat[]{
            \centering
            \includegraphics[width=.5\textwidth]{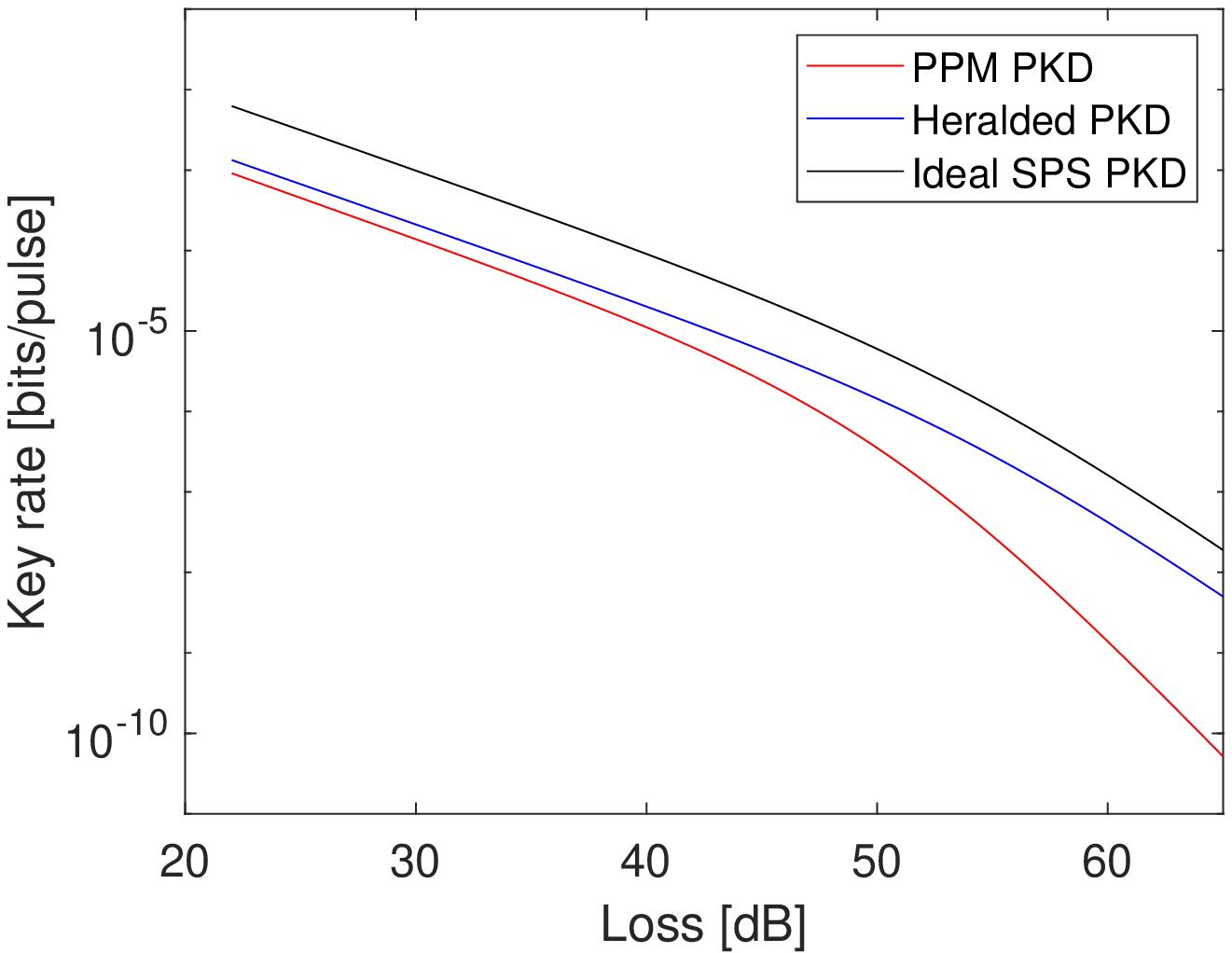}
            \label{fig:pkd_key}
        }

\subfloat[]{
             \centering
             \includegraphics[width=.5\textwidth]{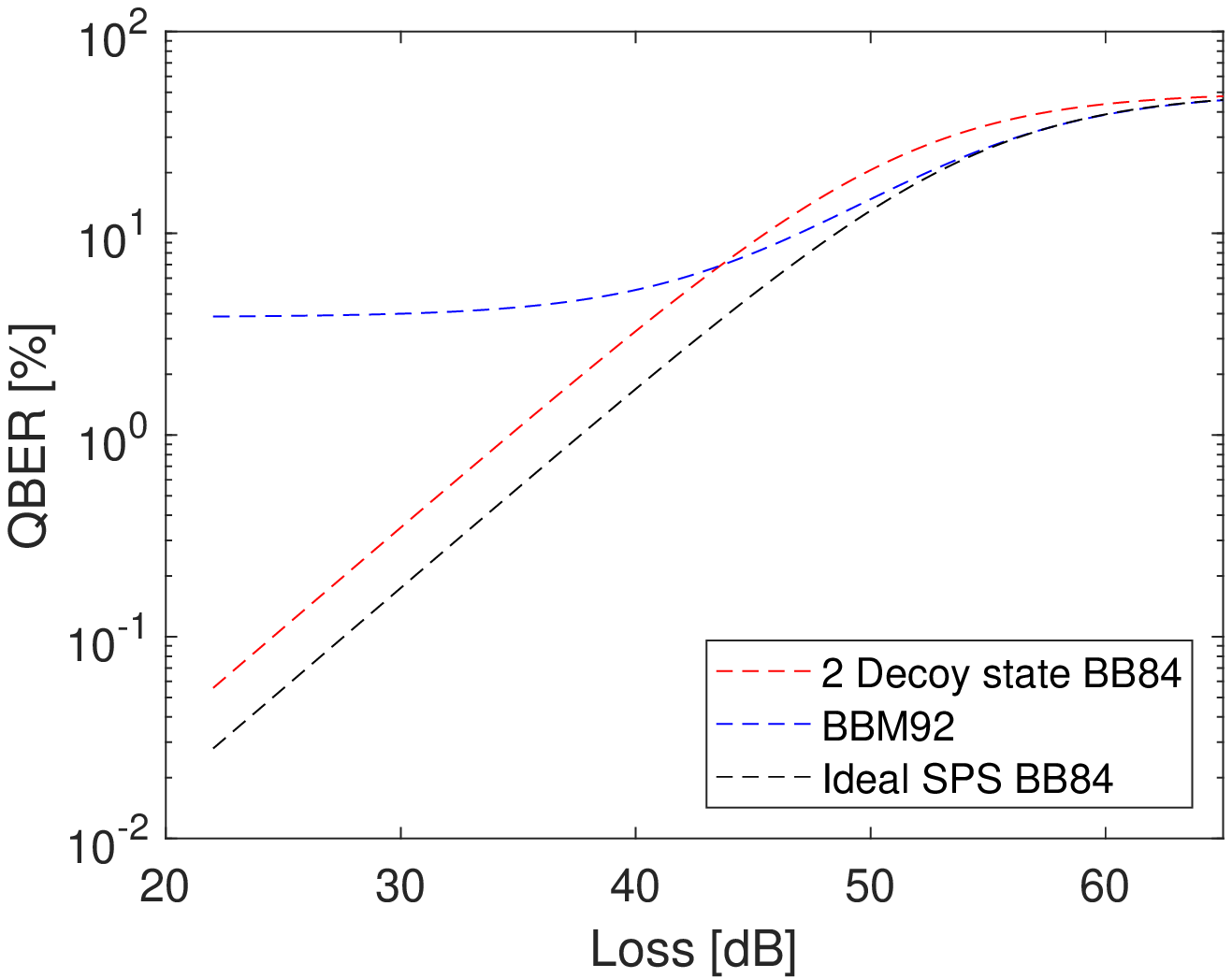}
             \label{fig:qkd_qber}
       }
\subfloat[]{
             \centering
             \includegraphics[width=.5\textwidth]{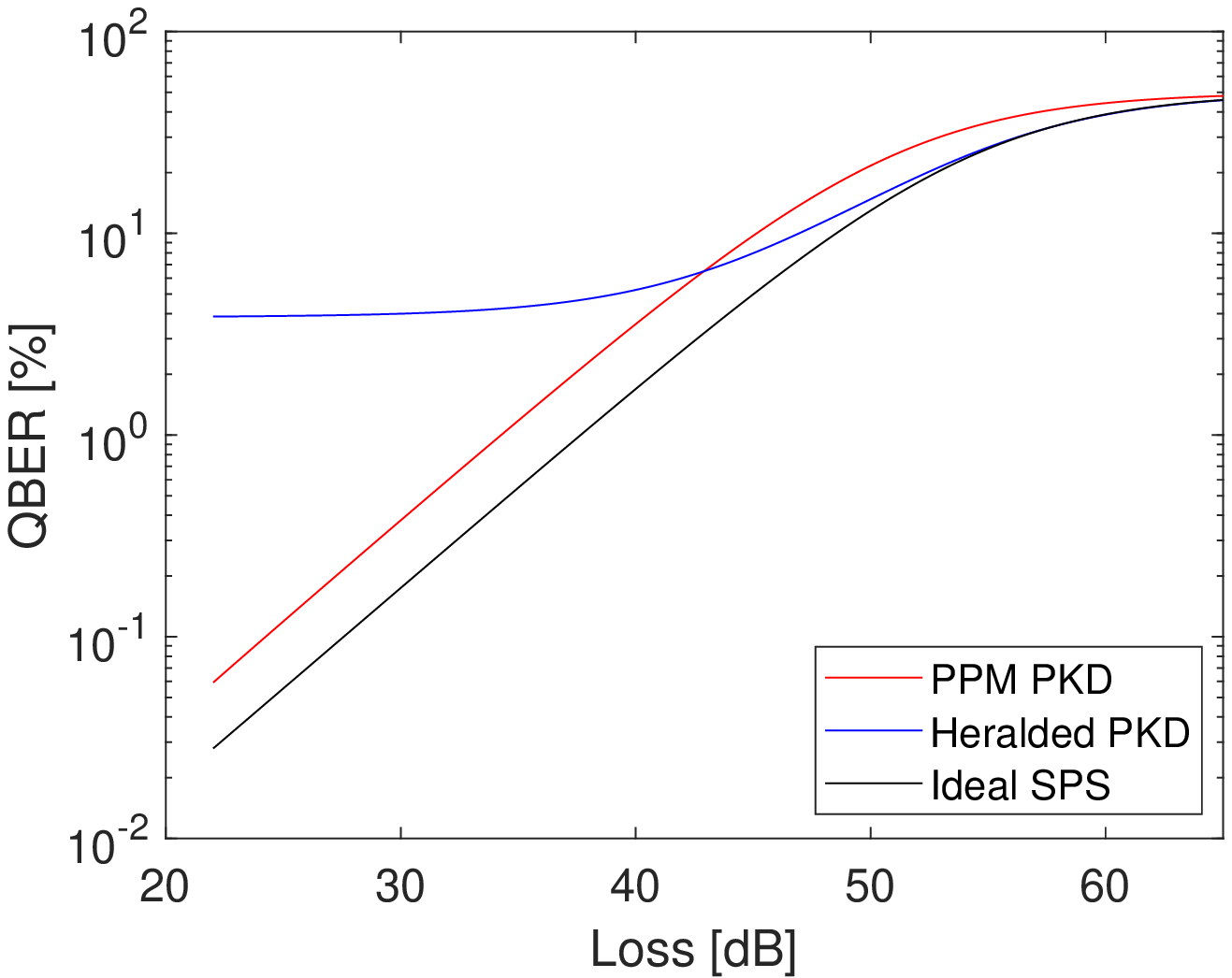}
              \label{fig:pkd_qber}
       }
\caption{Results of modelling the schemes described in Table~1.\\
(a) Key generation rate for Quantum Key Distribution (QKD) protocols: Decoy State BB84, BBM92, and BB84 using an ideal Single Photon Source (SPS). \\
(b) Key generation rate for photon key distribution (PKD) schemes: Pulse-position modulated (PPM) Photon Key Distribution (PKD), PKD with a heralded photon source, and PKD using an ideal SPS. \\
(c) Quantum Bit Error Rate (QBER) for QKD protocols. \\
(d) QBER for simplified PKD schemes. } \label{fig:results}
\end{figure*}







\section{Discussion}\label{sec:discussion}

The results are discussed in the sections below based on the different hardware types. General assumptions are that space-based detectors have 15,000 dark counts per second while ground-based detectors have 2,500 dark counts per second. This is based on pessimistic cases for single photon detectors before and after radiation damage \cite{Tan2013}. Furthermore, ground-based detectors are assumed to have an additional 1,000 background counts per second due to scattered light entering the receiver. This is slightly more conservative than noise counts estimated in recent literature \cite{Bourgoin2012}. All sources are assumed to have perfect visibility, e.g. for a SPDC source this means it is assumed to produce perfectly entangled states. 

\subsection{Weak Coherent Pulse (WCP) source}

Satellite-to-ground decoy state QKD using a Weak Coherent Pulse (WCP) source has conclusively been demonstrated on the Micius satellite in 2016 \cite{Liao2017a} and the QUBE mission aims to demonstrate a WCP QKD system on a nanosatellite \cite{Haber2018}. WCP sources are highly attenuated lasers which approximate a single photon state although multi-photon emissions are possible. We use the Micius source parameters as a realistic assumption in our model; on average our source emits 0.8 photons per pulse. For QKD it performs the decoy state BB84 protocol \cite{Ma2005} where it also emits decoy states of mean photon number 0.1, and operates at a ratio of 0.5: 0.25: 0.25 for signal, decoy and vacuum states. 

For WCP PKD no decoy states are transmitted, only signal states with an optimised mean photon number of between 0.4701 and 0.4583 for losses of 20-70dB. These are assumed to be coupled to a modulator that enables pulse position modulation (PPM) to encode bits as early (bit 0) or late pulses (bit 1). Compared to polarisation encoding this also removes the requirement to synchronise polarisation reference frames which is especially convenient for then communicating to satellites. 

\subsection{Spontaneous parametric down conversion (SPDC) photon pair source}

For the QKD protocols considered here the SPDC photon pair source is assumed to be an entangled photon pair source performing the BBM92 protocol \cite{Bennett1992,Naughton2019}. The Micius satellite demonstrated an entangled photon pair source in orbit \cite{Yin2017,Yin2017a} and the SpooQy-1 nanosatellite will demonstrate a miniaturised device in orbit in 2019 \cite{Grieve2018}. The PKD protocols with this setup can use the same hardware or a simpler correlated (heralded) photon pair source, with unit visibility, such as the miniaturised device flown on the Galassia nanosatellite \cite{Tang2016b}. To avoid synchronising reference frames the two-state encoding could be in left and right hand circular polarisation. 

Protocols using pair sources require two detections i.e. one of the pair must be detected at the source and the other of the pair detected at the receiver. The total detection efficiency at the source is assumed to be 25\% (i.e. there is a system loss of -6dB). The main effect of this approach is that the Quantum Bit Error Rate (QBER) partly becomes a function of the accidental rate of coincident detection between the two detectors. This accidentals rate is given by $S_1 \times S_2 \times \tau$ \cite{Janossy1944,Grieve2015a}, where $S_1$ and $S_2$ are the singles rates observed at the detectors, and $\tau$ the timing coincidence window. 

In this study, we assume that the source is capable of producing photon pairs at a raw rate of $1 \times 10^8$ per second into a single spatial mode, and that the timing resolution is $1 ns$, which are readily accessible \cite{Cao2018}. This leads to a QBER that is between 3\%  and 4\%; this QBER has a linear relationship with the timing window - faster electronics will lead to smaller QBER. An advantage of using photon pairs is that the probability of multi-photon emission is insignificant (similar to single photon sources), leading to a smaller percentage of the raw key being discarded in privacy amplification.

\subsection{Single photon source (SPS) }

The SPS is assumed to be a perfect device that produces single photons on demand and never multi-photon pulses. For QKD this means it can perform the BB84 \cite{Bennett2014} protocol without any requirement for decoy states. For PKD it can even be encoded in wavelength giving a broad scope for multiplexing. In both cases the SPS has the highest rates. It should be noted that compared to the other technologies SPS devices are currently at low `technology readiness levels' \cite{Vogl2019} or only work at cryogenic temperatures which means they are not a practical choice for deployment in space.

\subsection{Summary}

The PKD protocols have higher rates than the QKD protocols because eavesdroppers have no access to the channel and only `collective attacks' are allowed. That means if there are no multi-photon emissions (i.e. not using the WCP source) no privacy amplification is required. PPM PKD suffers from multi-photon emissions so requires privacy amplification. In general the relaxed security assumptions increase key rates and allow for key generation at larger distances. Comparing between different hardware methods of PKD (or between different hardware methods of QKD) is less conclusive, since the `pulse' rates of different implementations of the hardware will vary. 
\section{Materials and Methods}


The decoy state BB84 equations used are those published in reference \cite{Ma2005}, and the BBM92 equations are those from reference \cite{Ma2007a}. 	

The equation for PPM PKD is derived from the Devetak-Winter bound for collective attack for a simplified COW PPM scheme with decoys from appendix B of reference \cite{Branciard2007}. All parts due to decoy are set to neutral, and an error correction term is added to the equation. The final key rate is thus

\begin{equation}
 R(\mu_{opt}) = g(\zeta) \times \frac{t}{(1-t)} \times \eta \times (1 - H_{2}(QBER) \times f_E) \times f_{source}      
\end{equation}

where $g(\zeta)$ is a privacy amplification term corresponding to the optimised mean photon number $\mu_{opt} = \frac{\zeta}{1-t}$. $\eta$ is the detection efficiency, $f_{source}$ is the source frequency and $t$ is the transmissivity. 




\section{Conclusions}
The approach presented in this paper can be applied to key distribution, not only when using satellites, but also in other types of FSO-based systems. For example, in ad-hoc networks where the transmitter and receiver can visually identify each other, it may be straightforward to assess the absence of a man-in-the-middle and use PKD type protocols to establish a secret key. 
If the possibility of man-in-the-middle attacks can be excluded or monitored by conventional means, then PKD protocols can allow for secure key distribution using quantum communication methods at increased losses using similar or simplified hardware. In a practical key distribution scenario, it is worthwhile to have a better specification of the threat model and to devote efforts in closing physical side-channels, rather than attempting to defeat a hypothetical quantum adversary. Finally, we remark that using QKD hardware it may be possible to switch to a PKD protocol when the channel is assessed to be free of active attacks, and thus increase key distribution rates.


\vspace{6pt} 









\pagebreak

\textbf{Acknowledgements}\\
We thank Charles Lim and Ignatius William Primaatmaja for their valuable feedback and suggestions.\\

\textbf{Patents}\\
A patent has been filed for the heralded source key distribution method.\\

\textbf{Contributions} \\
Conceptualization: Robert Bedington, James A. Grieve and Alexander Ling; Data curation: Tom Vergoossen; Formal analysis: Tom Vergoossen; Funding acquisition: Alexander Ling; Investigation: Tom Vergoossen; Methodology: Tom Vergoossen; Project administration: Robert Bedington and Alexander Ling; Resources: Alexander Ling; Software: Tom Vergoossen; Supervision: Robert Bedington and Alexander Ling; Validation: Tom Vergoossen; Visualization: Tom Vergoossen; Writing – original draft: Tom Vergoossen and Robert Bedington; Writing – review and editing: Tom Vergoossen, Robert Bedington, James A. Grieve and Alexander Ling.\\



\textbf{Funding} \\
This work is partially supported by the National Research Foundation, Prime Minister’s Office, Singapore (under the Research Centres of Excellence programme and through Award No. NRF-CRP12-2013-02).\\

\textbf{Abbreviations}\\
The following abbreviations are used in this manuscript:\\
\begin{tabular}{@{}ll}
QKD & Quantum Key Distribution\\
FSO & Free Space Optics\\
PKD & Photon Key Distribution\\
SPS & Single Photon Source\\
WCP & Weak Coherent Pulse \\
PPM & Pulse Position Modulation \\
BB84 & Bennett Brassard 1984 (QKD protocol \cite{Bennett2014})\\
BBM92 & Bennett Brassard Mermin 1992 (QKD protocol \cite{Bennett1992})\\
AES & Advanced Encryption Standard\\
QBER & Quantum Bit Error Rate \\
COW & Coherent One Way (QKD protocol \cite{Branciard2007})\\
SKA & Secret Key Agreement \\
SPDC & Spontaneous Parametric Down Conversion
\end{tabular}
\\




%

\end{document}